**Running title:**

# Time-lag bias induced by unobserved heterogeneity: comparing treated patients to controls with a different start of follow-up

**van Eekelen R[1*], Bossuyt PMM[2], van Geloven N[3]**

[1] Epidemiology & Data Science, location VUmc, Amsterdam University Medical Centres, Meibergdreef 9 1105 AZ Amsterdam, the Netherlands

[2] Epidemiology & Data Science, location Academic Medical Centre, Amsterdam University Medical Centres, Meibergdreef 9 1105 AZ Amsterdam, the Netherlands

[3] Department of Biomedical Data Sciences, Leiden University Medical Centre, Einthovenweg 20, 2333 ZC Leiden, the Netherlands

* Correspondence address. Email: r.vaneekelen@amsterdamumc.nl

**Abstract**

In comparative effectiveness research, treated and control patients might have a different start of follow-up as treatment is often started later in the disease trajectory. This typically occurs when data from treated and controls are not collected within the same source. Only patients who did not yet experience the event of interest whilst in the control condition end up in the treatment data source. In case of unobserved heterogeneity, these treated patients will have a lower average risk than the controls. We illustrate how failing to account for this time-lag between treated and controls leads to bias in the estimated treatment effect. We define estimands and time axes, then explore five methods to adjust for this time-lag bias by utilising the time between diagnosis and treatment initiation in different ways. We conducted a simulation study to evaluate whether these methods reduce the bias and then applied the methods to a comparison between fertility patients treated with insemination and similar but untreated patients. We conclude that time-lag bias can be vast and that the time between diagnosis and treatment initiation should be taken into account in the analysis to respect the chronology of the disease and treatment trajectory.

**Keywords:** selection bias, time-lag bias, survivorship bias, heterogeneity, frailty, survival analysis, left truncation, Cox models, insemination, IVF

## Introduction

When observational data are used to estimate the effect of treatment in comparative effectiveness research, treated patients and controls could have a different start of follow-up.[1] We here define controls as those given no treatment, usual care, first-line treatment, or any other form of not providing the treatment of interest. Often, patients may remain untreated after a diagnosis has been made and initiate treatment later in their trajectory. Similarly, patients who received first-line treatment may end up needing second-line treatments later.[2] Differences in the start of follow-up are common when data on both conditions are not collected within the same source, such as when first-line care is not provided by the same hospital that provides the treatment.

Some outcomes, such as death, by definition prevent the patient from continuing their trajectory. Patient who experience such an outcome cannot receive the treatment of interest later.[3-11] This leads to differences between treated and control populations. An example is a study of ours in which we aimed to compare in vitro fertilisation (IVF) treatment to expectant management, which is usual care without treatment, in couples with unexplained infertility.[12] The source of treatment data was the UK national IVF registry and only covered patients from treatment initiation onwards. The source of control data was a Dutch prospective cohort that followed patients from diagnosis until the earliest of pregnancy or treatment initiation. Hence, the start of follow-up was different for these two data sources, which led to differences between treated patients and controls.

Differences between treated patients and controls can stem from two mechanisms: confounding and selection. For the purpose of this paper, we consider confounders to be known, measured and adjusted for in the analysis. A common definition of selection bias is the difference in the treatment effect found in the cohort selected for analysis and the treatment effect in the full population eligible for treatment.[6] When comparing treated and control patients with a different start of follow-up, a specific type of selection bias is caused by patient characteristics related to the outcome that are unbeknownst to the clinician and patient when deciding on treatment. Differences due to unmeasured characteristics are referred to as unobserved heterogeneity, or as frailty in survival analysis literature.[13] Unobserved heterogeneity leads to differences in outcomes between treated and untreated patients because only patients who did not yet experience the event of interest whilst in the control condition can end up in the treated data source. These patients will on average have a lower risk for the event than the controls. Patients in the two separate sources thus differ in their risk profiles, even after adjustment for all measured confounders.[11] A naïve comparison of treatment groups will lead to a biased treatment effect, as the comparison uses two misaligned time axes. We refer to this as time-lag bias induced by unobserved heterogeneity.[2,14]

Other issues related to misalignment of time axes are immortal time bias and the depletion of susceptibles over follow-up time.[2-9] The first, immortal time bias, is a self-inflicted error during the analysis when conditioning on future information, such as assigning a patient to the treatment group at diagnosis when in reality they received treatment later.[2,8,9] With 'depletion of susceptibles', we mean that the hazard ratio attenuates to the null over time as risk sets become increasingly similar over time, even if there is a true non-null treatment effect.[5] Both issues differ from the time-lag bias we discuss here. In the setting of time-lag bias due to unobserved heterogeneity, bias is already present at inclusion due to the nature of the two data sources and their start of follow-up.

Azzato *et al.* (2009) and Eijkemans *et al*. (2019) suggested selecting only patients immediately after they become eligible for treatment, so that no heterogeneity-induced selection could have taken place.[10,15] This implicitly suggests that a variant of the average treatment effect in the controls (ATC) is targeted, which may or may not be the estimand of interest. There are other issues with this approach: it greatly reduces sample size or might be even impossible due to non-positivity, e.g., if the clinical protocol excludes patients from receiving treatment shortly after diagnosis.

Here we explore five approaches to account for time-lag bias when comparing treated and control patients. All approaches utilise the time between diagnosis and treatment start, which we refer to as *waiting time* from now on. In simulations and in an application to clinical data, we compare the following methods: selecting treated who started treatment immediately, selecting controls who survived up to the median time of treatment initiation, left truncation of waiting time, adding waiting time as a covariate and sequential landmarking.

## Methods

### *Notation*

We wish to estimate the causal effect of treatment $A$ ($a = 1$) relative to control ($a = 0$) on time to some event $Y$ for patients with a certain disease.

After diagnosis, not all patients may receive treatment immediately: more severe cases might start earlier than the less severe cases. Here, disease severity can serve as an example of a measured confounder, denoted by $F$ in the directed acyclic graph (DAG) in **Figure 1A**, where the relationships between $F$, treatment A and event status at time $k$ and time $k + 1$ are depicted.

We can adjust for $F$ as usual in observational data analysis, for instance by regression analysis, matching or inverse probability weighting, depending on the estimand of interest.[16] If only $F$ is of influence and accounted for, inference is valid both for the effect of $A$ on survival between time zero and a certain time $k$ and for the effect on survival from $k$ onwards. Now consider a factor that leads to a higher probability of survival, for instance a gene combination $G$, which is unknown to clinicians and patients, therefore generally unmeasured and also independent of $F$. As $G = 1$ is related to survival up to time $k$ but not related to $A$, $G$ is not a confounder at diagnosis as shown in **Figure 1B**. As time progresses, patients with $G = 1$ are more likely to have survived. Because treatment is started later, the proportion of treated patients with $G = 1$ increases over time, thus introducing an association between $G$ and $A$. Conditioning on the collider survival up to time $k$, which we implicitly do in the mentioned study design involving a data source with treated patients later on, now opens up a path between $G$ and $A$ that biases the causal association between $A$ and survival from $t$ onwards, as shown in **Figure 1C.**

**Figure 1**. Directed acyclic graphs to show the influence of confounders (*F*) and unobserved risk factor(s) (*G*) in the causal pathway between treatment (*A*) and the two outcomes, event status at a certain time *k* ($Y_k$) and event status at a later time point *k* + 1 ($Y_{k+1}$). In graph **A.**, there is no factor *G*. If *F* comprises a sufficient adjustment set and is conditioned on in the analysis, the effect of *A* on $Y_k$ and $Y_{k+1}$ can be estimated without confounding bias. In **B**. at time zero i.e. diagnosis, *G* is related to $Y_k$ but not to *A*, thus *G* is not a confounder. At later timepoints shown in **C**., when conditioning on the collider survival up to time *k* ($Y_k$), this opens a path between *G* and *A* that biases the effect of *A* on $Y_{k+1}$. A box around a variable indicates conditioning on that variable.

**A**. without *G*

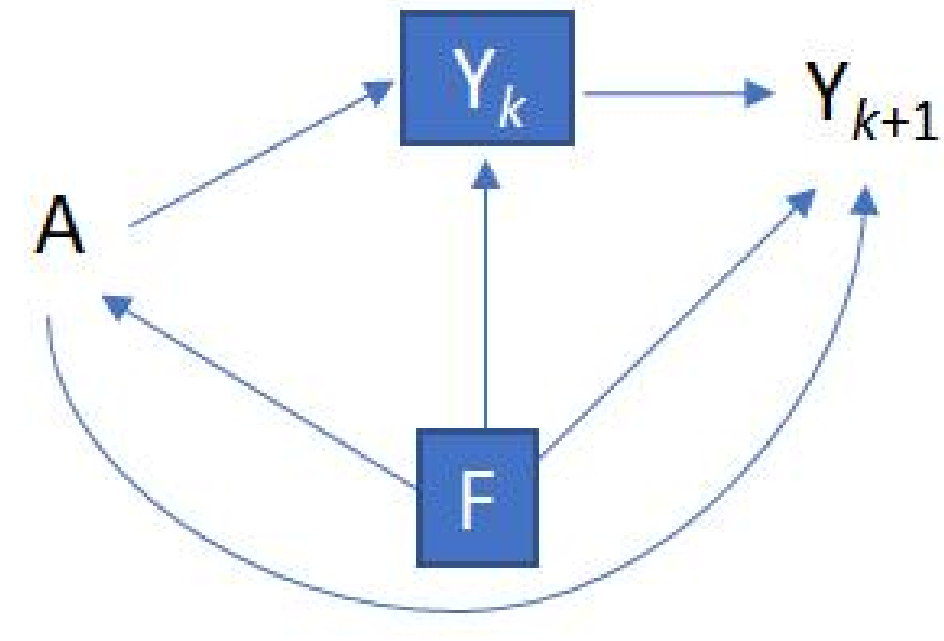

**B**. with *G*, at time zero
i.e. diagnosis

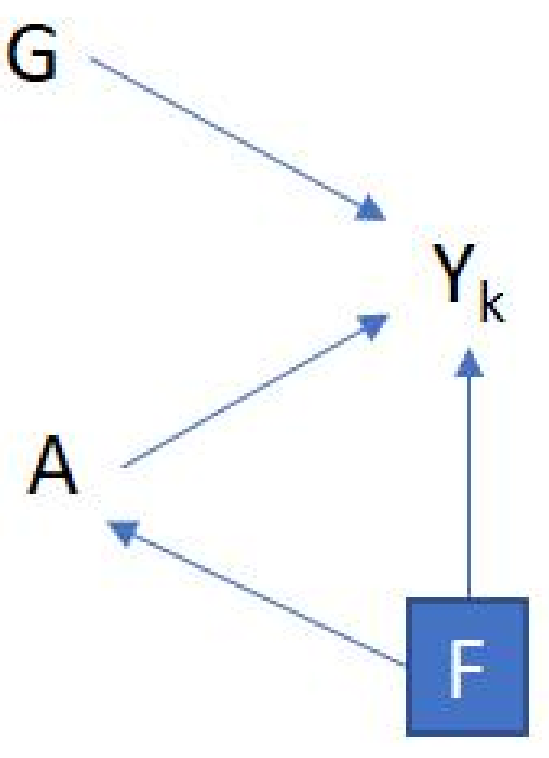

**C**. with *G*, at later
timepoint *k*

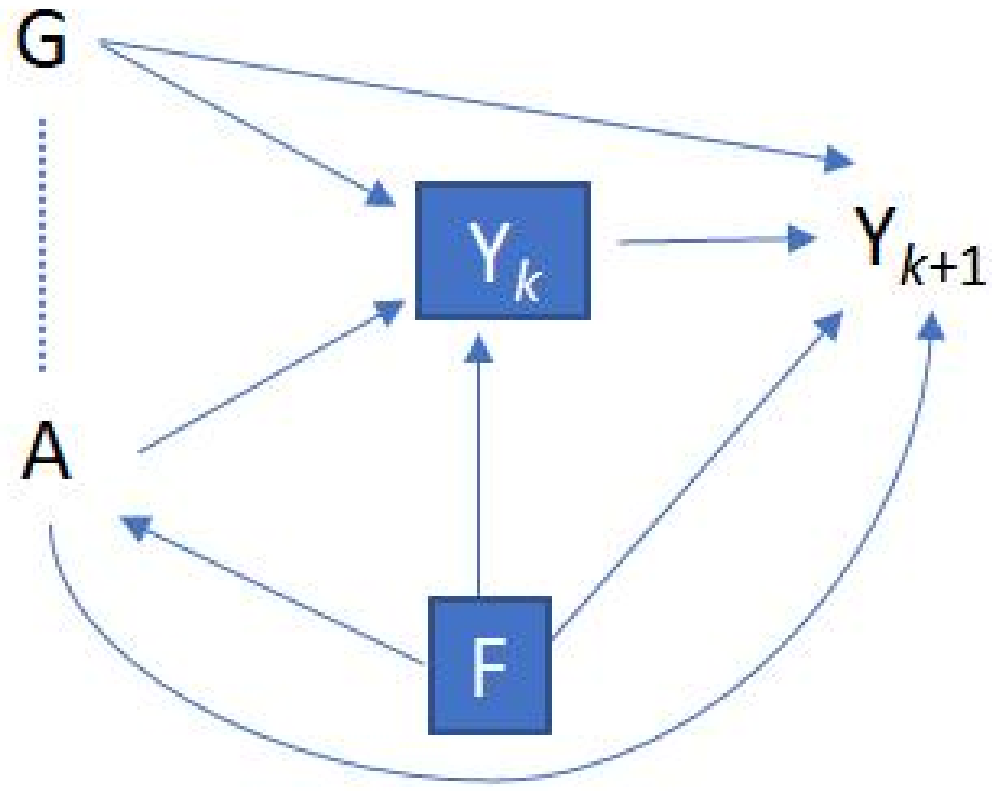

*Estimands and timelines*

We distinguish between the true conditional treatment effect, accounting for heterogeneity, and the true marginal treatment effect which, due to attenuation of the hazard ratio towards the null during follow-up, depends on the time window of follow-up.

We follow notation as in Young *et al.*, with *Y* denoting the event indicator, i.e., failure, *k* = 0,1, … *K* follow-up time with *k* = 0 time of diagnosis, *A* treatment with *a* = 0 the control condition, and *a* = 1 the active treatment which we for now assume starts at *k* = 0 and is sustained until *K*. Then the counterfactual risk of *Y* by *k* + 1 had all patients been assigned to *a* can be defined as

$$\Pr\,[Y_{k+1}^{a} = 1] \tag{1}$$

such that the average causal effect of treatment A by *k* + 1 can be expressed as

$$\Pr\left[Y_{k+1}^{a=1} = 1\right] - \Pr\left[Y_{k+1}^{a=0} = 1\right] \tag{2}$$

In case of (censored) time-to-event outcome data, the effect of interest is commonly the hazard ratio.

The hazard at *k* + 1 of *Y* under *a* is

$$\Pr\,[Y_{k+1}^{a} = 1 | Y_{k}^{a} = 0] \tag{3}$$

which would give the average causal hazard ratio for *A* as

$$\Pr\left[Y_{k+1}^{a=1} = 1 | Y_{k}^{a=1} = 0\right] / \Pr\left[Y_{k+1}^{a=0} = 1 | Y_{k}^{a=0} = 0\right] \tag{4}$$

As the contrast in counterfactual hazards is conditional on survival to *k*, it is only interpretable as a true causal effect if and only if there are no differences between patients other than *A*, an assumption which never holds due to natural differences between patients. In the presence of unobserved heterogeneity, the relative effect attenuates towards the null. In other words, if there are prognostic differences between patients that are unaccounted for or unmeasured, the causal effect in interval *k* to *k* + 1 will differ from the causal effect in interval *k* +1 to *k* + 2. The larger *K* gets and the more heterogeneity there is, the more the last interval will differ from the first, which is known as depletion of susceptibles.[17]

As heterogeneity is practically never fully explained by measured covariates, a true ‘causal’ hazard ratio conditional on heterogeneity is unattainable. We choose as the effect of interest the marginal hazard ratio averaged over a defined range of follow-up in the data, i.e., up to *K*.

We denote the time after diagnosis at which treated patients started their treatment as *T* and the counterfactual outcome Y at time k + 1, if a patient would be treated at time point *t* (and remain treated thereafter) as $Y_{k+1}^{t}$. We then define an ‘ATC’ type of estimand as for those eligible for treatment who did not start treatment at 0,1,… *K*, what if they would have started immediately i.e. *t* = 0. We get this ATC estimand by averaging the following hazards over the range 0,1,…,*K*.

$$\frac{\Pr\left[Y_{k+1}^{t=0} = 1|T > K, Y_{k}^{t=0} = 0\right]}{\Pr\left[Y_{k+1}^{t>K} = 1|T > K, Y_{k}^{t>K} = 0\right]} \tag{5}$$

We define an ‘average treatment effect in the treated (ATT) type of estimand as for those eligible for treatment who did start treatment before *K*, what if they would have never started treatment (or started after *K*). We get the ATT by averaging the hazard over time range 0,1,…,*K* and over all observed treatment waiting times:

$$\frac{\Pr\left[Y_{k+1}^{t=t^{*}\in T} = 1|T < K, T > k, Y_{k}^{t=t^{*}\in T} = 0\right]}{\Pr\left[Y_{k+1}^{t>K} = 1|T < K, T > k, Y_{k}^{t>K} = 0\right]} \tag{6}$$

With $t^{*} \in T$ being all observed values of treatment start *t* between 0 and *K*.

The ‘average treatment effect (ATE)’ is defined as the weighted average of (5) and (6) i.e., the combined effect when all those eligible for treatment that did not start treatment would have started treatment at *t* = 0 and those eligible for treatment who did start treatment would have never started treatment.

*Estimation approaches*

To show what problems may arise when using ad-hoc estimation strategies to combine treated and control data sources and how this may lead to targeting a different estimand to those specified above, we visualise different estimation strategies in **Figure 2.**

**Figure 2. Blue lines representing follow-up for control and red lines follow-up for treated patients. Black arrows represent treatment contrasts. Dotted lines represent left truncation until the start of follow-up.**

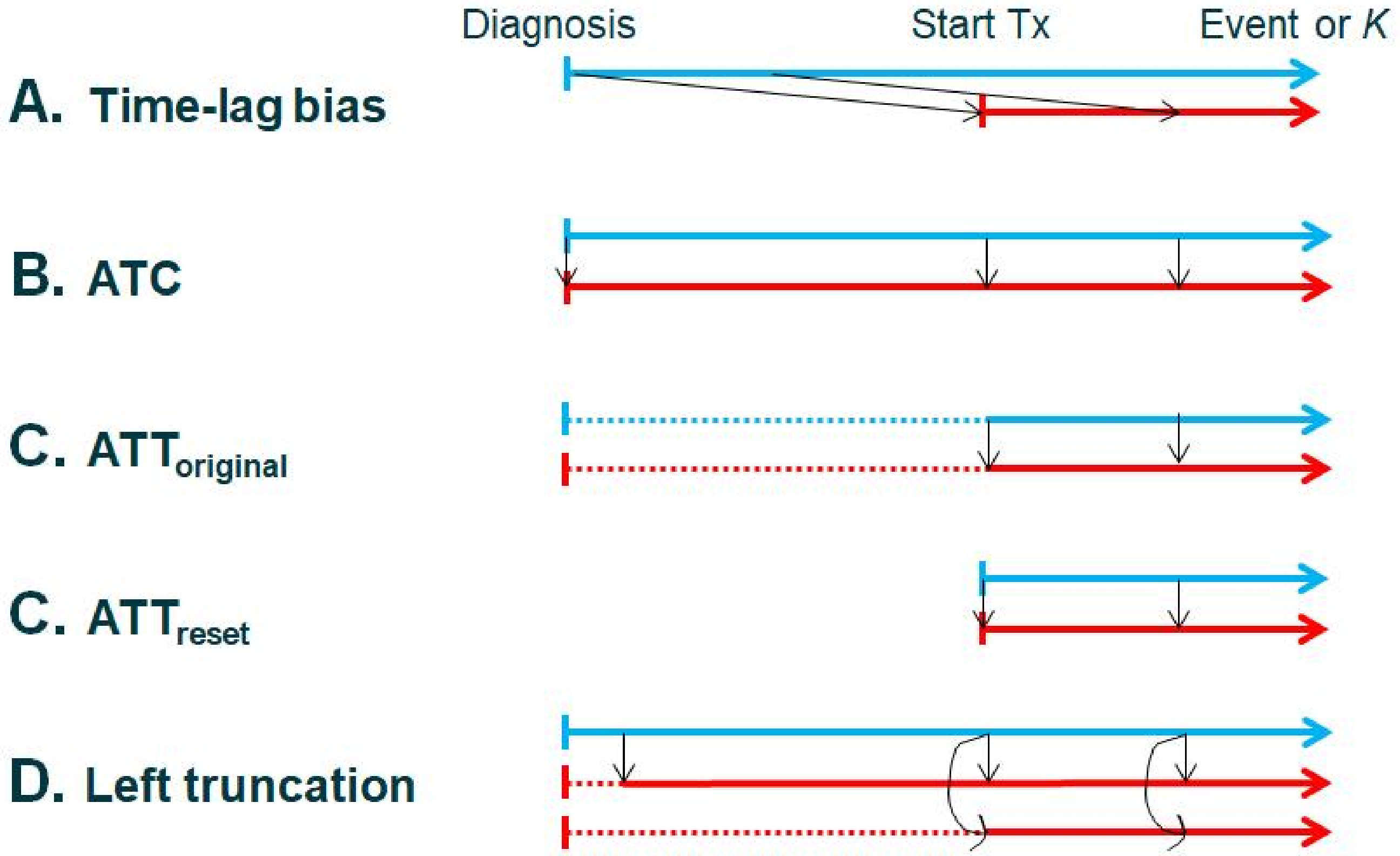

**Misalignment of start of follow-up for control and treated individuals.**

An estimation strategy which is sometimes used erroneously in practice is to consider starting time of treatment (*T*) as the start of follow-up (i.e., time zero) for the treated patients whereas for the control patients the starting time is diagnosis, shown in **Figure 2A**. Using *T* as time zero 'resets' the follow-up time of treated patients from treatment onwards, and is effectively deducting *t* from *k*. This induces bias as the comparison is no longer fair and actually targets a different estimand which can be expressed as

$$\frac{\Pr\left[Y_{k+1+t}^{t=t^{*}\in T}=1 | Y_{k+t}^{t=t^{*}\in T}=0\right]}{\Pr\left[Y_{k+1}^{t>K}=1 | Y_{k}^{t>K}=0\right]} \tag{7}$$

As is visible from (7), follow-up is misaligned between treated patients and control patients, with a comparison at time $k + 1$ for control to time $k + 1 + t$ for treated. This is the situation in which time-lag bias occurs.

In the following estimation approaches, the issue of not having observed the treated patients from $k = 0$ onwards is accounted for in different ways, either by selection of patients that survived up to a certain time point or by conditioning on $T$.

**Selecting treated patients who started immediately after diagnosis.**

In **Figure 2B** the time axes are aligned by selecting follow-up from treated patients who started treatment (almost) immediately after diagnosis, given a certain grace period of choice. For instance, within one month as the time unit would give $0<t<1$. This strategy targets the ATC estimand in (5). The effect is only interpretable for a subgroup of all patients: the controls in the observed data. It also requires that there are sufficient early treated patients available.

**Selecting controls who survived up to a certain treatment start, keeping the original time axis.**

In **Figure 2C ATT (original time)**, time-lag bias is prevented by comparing treated to selected controls who survived the same period. This approach targets the ATT estimand described by (6). This is complicated to estimate as there are many $t$ and the effect is only interpretable for the observed pattern of treatment starts $T$.[18] The time axis is here referred to as ‘original time’ because the perspective is from defining time zero as the time of diagnosis.

**Selecting controls who survived up to the median treatment start, redefining time zero as treatment starts.**

For ATT, we could also change the perspective of the time axis to by setting the treatment starting time $t$ as time zero, shown in **Figure 2C ATT (reset time**). The target estimand can then be described as the averaging the following hazard over $k$ = 0,1,…$K$ and over all observed treatment starts:

$$\frac{\Pr\left[Y_{k+1+t}^{t=t^* \in T} = 1 | T = t^* < K, Y_{k+t}^{t=t^* \in T} = 0\right]}{\Pr\left[Y_{k+1+t}^{t>K} = 1 | T = t^* < K, Y_{k+t}^{t>K} = 0\right]} \tag{8}$$

Treated patients are again compared to controls surviving the same period, but now with a time reset to zero at treatment start onwards for both. This differs from the previous estimand because all treated patients are combined as if all starting follow up at the same time point. This strategy again aligns the time axes, thereby circumventing time-lag bias but the effect is only interpretable for the observed pattern of treatment starting times.

As mentioned, as the effect on the original time axis may be complicated to estimate due to the range of $t$, one could consider a simpler approach: selecting all controls who survived up to the median $t$ and following them onwards.[19] There needs to be a sufficient number of controls who survived up to that point and residual time-lag bias remains as there is still variation in $t$ for treated patients.

**Left truncation.**

The time axis in **Figure 2D** follows the combination of (5) and (6) on the original time axis and is thus defined as the ATE as it combines the ATT and ATC, using all data. The intuitive thing here is to consider the waiting time $t$ as the time a treated patient 'enters' the survival analysis risk set: left truncation, also referred to as delayed entry. This is intuitively modelled in Cox regression using left truncation for $t$, as that is the time when treated patients enter and contribute information. The definition of the model risk set and likelihood contributions change such that a treated patient with waiting time $t$ is only compared to controls that survived up to $t$ as well, following **2D**.

**Adding waiting time as covariate and repeated landmarking.**

There are two alternatives for estimating the ATE which can also be conducted on the reset time axis.

Regressing on waiting time, i.e., adding waiting time *t* as a covariate to the outcome model, is fairly straightforward and is also flexible, as it can be used in any analysis model that can take continuous covariates. The functional form of *t* in the model might be crucial for valid inference and is not necessarily linear. We suggest pre-specifying forms based on field-specific empirical evidence. If that is unavailable, the fit of a linear function should be compared to more complex functions, e.g., a quadratic function or, if sample size allows, (restricted) cubic splines.[20]

A second approach to estimating the ATE is what is known as landmarking in survival literature: constructing multiple datasets for sequential time points selecting treated patients and controls still at risk. This can be viewed as 'matching' on *t* (instead of matching on patient characteristics). Thus, we select patients who started treatment at *t* and controls that survived up to *t* as well. Because *t* varies, and we wish to use all data for ATE, this has to be done repeatedly for all *t.* We have to choose the sequential time points and the caliper or grace period i.e. how 'early' treatment needs to be started in order to be grouped together.[21] For instance, landmark datasets can be constructed for sequential time points $w = \{0,1,2 \ldots K\}$ such that treatment has to be started within 1 time unit after the start of the landmark *w*. All these landmark datasets can be pooled in analysis with e.g. a Cox model, stratifying on the dataset identifier *w*.[21,22]

When using all data from treated and control patients, without targeting a pre-specified group, this targets ATE.

As controls may appear in the pooled data multiple times, robust (sandwich) standard errors that take this correlation into account are necessary.

**Simulation**

We assessed whether these methods reduce time-lag bias in a simulation study.

*Simulation setup*

Our simulation setup is loosely based on our work in the field of fertility treatments.[23] Using previous notation, $a = 1$ is used for treatment and $a = 0$ for control, $Y$ the event indicator with $y = 1$ representing the event occurred and follow-up time $k$ with $k = 0,1,\ldots K$. We assume that the event is favourable for patients, such as recovery or pregnancy from our motivating example, and that treatment $a = 1$ increases the probability of recovery. We simulate a large sample size as we are interested in structural bias and not in sampling variation.

We generated two cohorts: one for the control patients and one from which we selected the treated patients. For both, the cohort starts with n=5,000,000 patients included at diagnosis. For all patients, we drew an untreated time to event from an exponential distribution using individual hazard rates generated via two mechanisms that introduce unobserved heterogeneity: first via a beta distribution for the probability of the event which was then transformed to a hazard rate, and second via a proportion of patients with a lower risk of the event due to an unobserved binary risk factor $G$ (details and parameters in the **Appendix**).

All patients from the first cohort were selected to be used as controls in the simulation. For selecting treated patients from the second, we additionally drew time to start of treatment, i.e., waiting time $t$, using a second exponential distribution with hazard rate 0.1. We then selected the patients from the second cohort who did not recover before $t$, as only those non-recovered patients would be treated. We applied a time-fixed treatment hazard ratio (conditional on heterogeneity) to multiply the individual patient's untreated hazard rate. Then we used those treated hazard rates to draw the treated time to event from a third exponential distribution.[24] We combined the treated and control cohorts in one dataset and applied a naïve method to estimate the treatment effect in terms of a hazard ratio plus the five proposed adjustment methods to account for time-lag bias due to heterogeneity. No censoring was applied.

*True values*

The treatment effect applied during simulation of the data is the true hazard ratio conditional on heterogeneity. As explained earlier, none of the methods will be able to estimate this hazard ratio due to attenuation of the hazard ratio towards the null over follow-up. Instead, we used as true values the hazard ratios marginalised over follow up time, which we derived from simulated counterfactual cohorts. We simulated these counterfactual cohorts by creating clones (i.e. same treated or untreated individual hazard rate and same time to treatment $t$) for patients included in either cohort following **Figure 2B** for ATC and **2C**$_{\mathbf{original}}$ for ATT, switching their treatment status and changing the individual hazard rate accordingly, then using said rate to draw another time to event under the alternative treatment status. For ATE, we combined the ATC and ATT counterfactual cohorts. True marginal hazard ratios were then estimated with Cox regression using $t$ as the time of left truncation.

*Comparison of methods*

We estimated hazard ratios from the simulated observational cohorts using Cox proportional hazards models, first unadjusted and then following the five methods described before: (1) selecting treated who started treatment almost immediately ($0<t<1$), (2) selecting controls at the median $t$, (3) using $t$ as the time of left truncation, (4) regressing on $t$ with varying functional forms and (5) landmarking using sequential time points *{0,1,2…20}*, with a grace period of 1 time unit and all 21 landmark datasets followed until the unrestricted maximum follow-up $K$.

**Table 1. Simulation results, n=5,000,000 per cohort.**

| Effect estimator | Scenario 1.<br>Beta heterogeneity<br>Hazard ratio: | Scenario 2.<br>Binary *G* factor<br>Hazard ratio: |
|---|---|---|
| **True*** | **1.500 (ATE)**<br>**1.490 (ATC)**<br>**1.510 (ATT)** | **1.500 (ATE)**<br>**1.485 (ATC)**<br>**1.525 (ATT)** |
| **Unadjusted** | 1.11 | 1.10 |
| **1. Select treated with $t \leq 1$ (ATC)** | 1.47 | 1.49 |
| **2. Select controls that survived up to median *t* (ATT)** | 1.39 | 1.32 |
| **3. Left truncation (ATE)** | 1.55 | 1.55 |
| **4. Waiting time as a linear covariate (ATE)** | 1.37 | 1.39 |
| **4. Waiting time as a quadratic covariate (ATE)** | 1.43 | 1.48 |
| **4. Waiting time as a covariate modelled with restricted cubic splines (ATE)** | 1.53 | 1.53 |
| **5. Sequential landmarking (ATE)** | 1.49 | 1.48 |

*t*, waiting time i.e. time between diagnosis and the start of treatment

* as estimated by Cox regression in counterfactual cohorts. Note that the true hazard ratio conditional on heterogeneity was 2.060 for Scenario 1 and 1.880 for Scenario 2.

*Simulation results*

In the scenario with beta heterogeneity, the unadjusted estimate for the hazard ratio (HR) was 1.11, erroneously indicating a much less effective treatment than the true marginal HR for ATE of 1.50 (**Table 1**). This bias is towards a relative effect of zero, so effective treatments with positive outcomes appear less effective and for negative outcomes it will appear as even more effective. This is because the selection can only lead to treated patients with on average a lower risk of the event than untreated patients. The estimates using left truncation (HR: 1.55), waiting time modelled as a covariate with restricted cubic splines (HR: 1.53), sequential landmarking (HR: 1.49) and selecting treated who started treatment immediately (HR: 1.47) were the closest to their true values.

Results for the scenario with the *G* factor were similar (**Table 1**).

## Application

For our application in clinical data, we used data on a prospective cohort in fertility patients, collected in seven centres in the Netherlands.[25-28] Patients in the cohort were followed for time to conception during a control period without treatment (expectant management, keep trying to conceive naturally) and, for those that did, data on the same patients after receiving treatment (intrauterine insemination).[23,25] As this was a prospective cohort, all patients were included at diagnosis. The treatment effect can then be estimated without time-lag bias using a time-varying covariate for insemination treatment in a Cox proportional hazards model: we consider this the 'truth' in this application.[9]

Next, to mimic the situation of comparing separate data sources, we purposefully split up the prospective cohort into a separate treated and control cohort and 'reset' the time axis for the treated, thereby starting follow-up for treated couples at their treatment initiation. Data for control couples remained unchanged but note that follow-up whilst in the control condition for those eventually treated were added to the control cohort to maintain the same population risk sets as in the analysis using the time-varying treatment covariate.

We first erroneously fitted a Cox model with a (time-fixed) treatment covariate that ignored the time-lag. We then applied the previously described five approaches using Cox models: selecting treated patients who started treatment within one month, selecting controls who survived up to the median time to treatment $t$ of 6.5 months, $t$ as a covariate with the restricted cubic splines fitted using 3 knots, and sequential time points *{0,2,4…16}* (in months, grace period of 2 months) all followed until $K$ = 18 months for sequential landmarking. In all analyses, we adjusted for the same set of confounders and used a robust sandwich variance estimator to adjust for the fact some couples appeared twice in the analysis.[29]

*Software*

Both the simulation study and the application were performed using R version 3.6.0 using the *survival*, *rms*, and *xtable* packages.[30]

## Data application results

*Descriptives*

Data on 1,896 unexplained subfertile couples were available, of which 800 received at least one insemination cycle within 18 months after inclusion, which was the cut-off for follow-up. 386 couples conceived naturally and 142 couples conceived after insemination. Median follow-up was 7 months for controls and 4 months for those eventually starting treatment. Several baseline characteristics for the two groups are summarized elsewhere.[28]

*Treatment effects*

**Table 2. Results from the application to data from fertility patients.**

| Effect estimator | Hazard ratio for insemination (95%CI): |
|---|---|
| **Time-varying covariate in the prospective cohort (the 'truth') (ATE)** | 2.15 (1.71-2.69) |
| **Unadjusted** | 1.62 (1.33-1.98) |
| **1. Select treated with $t \leq 1$ (ATC)** | 1.66 (1.16-2.39) |
| **2. Select controls that survived up to median $t$ (ATT)** | 2.04 (1.58-2.64) |
| **3. Left truncation (ATE)** | Here equivalent to time-varying covariate |
| **4. Waiting time as a linear covariate (ATE)** | 1.74 (1.27-2.38) |
| **4. Waiting time as a quadratic covariate (ATE)** | 2.11 (1.45-3.07) |
| **4. Waiting time covariate modelled with restricted cubic splines (ATE)** | 2.08 (1.43-3.01) |
| **5. Sequential landmarking (ATE)** | 1.98 (1.59-2.47) |

$t$, waiting time i.e. time between diagnosis and the start of treatment

+ Original axis refers to diagnosis as time zero for both cohorts. Reset axis refers to start of treatment as time zero for both cohorts. Time-lag axis refers to diagnosis as time zero for controls and start of treatment as time zero for treated patients.

Table 2 denotes the estimated treatment effects and their 95%CIs. For the Cox model with treatment as a time-varying covariate we found a 'true' hazard ratio (HR) of 2.15 (95%CI: 1.71-2.69) for insemination compared to no treatment.

After restructuring the data to re-enact the situation where two separate data sources are being compared with a different start of follow-up, we found an unadjusted HR of 1.62 (95%CI: 1.33-1.98): an underestimation of approximately 36% on log-hazard scale due to time-lag bias. Selection of treated couples who started treatment immediately gave a HR of 1.66 (1.16-2.39) and selection of control couples who did not conceive naturally for 6.5 months gave a HR of 2.04 (1.58-2.64). Using waiting time as a quadratic covariate yielded a HR for treatment of 2.11 (1.45-3.07), thereby eliminating 93% of the bias on log-hazard scale. Finally, sequential landmarking gave a HR of 1.98 (1.59-2.47).

## Discussion

We presented a prototypical setting in which treated patients and controls have a different start of follow-up. As a consequence, estimates of the causal effect of treatment may suffer from time-lag bias due to unobserved heterogeneity. This bias is towards a relative effect of zero and is already present at the start of follow-up, thereby different from attenuation of the hazard ratio towards the null during follow-up or immortal time bias.

We evaluated in simulations whether five estimation approaches could adjust for the time-lag bias. All estimators reduced time-lag bias to some extent but we found that three showed lowest bias: left truncation, sequential landmarking, and regressing on waiting time modelled with restricted cubic splines. In our application to a clinical study on insemination, we showed that even a simple procedure, including waiting time as a quadratic covariate in the model, eliminated the vast majority of time-lag bias, up to 93%.

Our approaches are not free of assumptions: for all, there needs to be overlap between controls still in follow-up and patients who started treatment, similar to the positivity assumption in confounding adjustment. For instance, this could mean that if for practical reasons or due to a certain treatment protocol there is always a long waiting time before treatment is initiated, it is not possible to estimate the ATC when this is defined as starting treatment at diagnosis.

In the simulation, we found that left truncation and regressing on waiting time using restricted cubic splines yielded slightly higher HRs than the true marginal HR, which was unexpected. We believe this is due to slight differences in the range of follow-up used for analysis, a result of defining risk sets using the 'reset' or 'original' time axis, which reduces the attenuation of the HR, albeit slightly.

All approaches require the waiting time between diagnosis and treatment initiation to be measured. If the waiting time is unmeasured, perhaps this can be estimated based on treatment protocols at the time. Lastly, we assume that there will always be some degree of heterogeneity that cannot be explained because it is due to unknown and/or unmeasured factors. We believe this is a reasonable assumption.

Example situations in which the methods in this paper can be useful to reduce bias due to heterogeneity include observational research on chronic diseases where treatments are not started directly at diagnosis or second-line treatments are only started in certain patients that survived long enough (e.g. rheumatology, nephrology, oncology).[2,14,31] Our approach also ties in with the discussion around preferring 'new-user' designs over 'prevalent-user' designs as well as the framework of target trial emulation, as it can prove difficult to define a common time zero for both treated and controls.[2,8,31-33] We showed that a simple, pragmatic procedure such as selecting controls at the median or average time of treatment initiation, used by Hernan *et al*. (2008) in an attempt to emulate a target trial, performed relatively poor in our simulations, although that same method eliminated the vast majority of bias in our application.[19]

In conclusion, observational research in which treated and control patients have a different start of follow-up may suffer from time-lag bias due to heterogeneity. This bias leans towards a relative effect of zero and is different from immortal time bias and from attenuation of the hazard ratio towards the null during follow-up. A uniform approach for recruitment or inclusion would work best, such as a prospective cohort recruited at time of diagnosis. Otherwise, the waiting time between diagnosis and treatment initiation should be measured routinely and used in the analysis to respect the chronology of the disease trajectory. Adding waiting time as a covariate in the analysis model is a straightforward way to deal with this bias although using waiting time as the time of entry through left truncation is also a good option when using a survival analysis model.

## References:

**1.** Rothman K, Greenland S, Lash TL. *Modern Epidemiology (4th edition)*. Philadelphia, PA: Lippincott Williams & Wilkins; 2012. p.

**2.** Suissa S and Azoulay L. Metformin and the risk of cancer: time-related biases in observational studies. *Diabetes Care* 2012;35:2665-73.

**3.** Aalen OO, Borgan O, Gjessing HK. *Survival and event history analysis: a process point of view.* New York: Springer; 2008. p.

**4.** Aalen OO, Cook RJ, Roysland K. Does Cox analysis of a randomized survival study yield a causal treatment effect? *Lifetime Data Anal* 2015;21:579-93.

**5.** Hernan MA. The hazards of hazard ratios. *Epidemiology* 2010;21:13-5.

**6.** Hernan MA, Hernandez-Diaz S, Robins JM. A structural approach to selection bias. *Epidemiology* 2004;15:615-25.

**7.** Keiding N, Albertsen KL, Rytgaard HC, Sorensen AL. Prevalent cohort studies and unobserved heterogeneity. *Lifetime Data Anal* 2019.

**8.** Suissa S. Effectiveness of inhaled corticosteroids in chronic obstructive pulmonary disease: immortal time bias in observational studies. *Am J Respir Crit Care Med* 2003;168:49-53.

**9.** Zhou Z, Rahme E, Abrahamowicz M, Pilote L. Survival bias associated with time-to-treatment initiation in drug effectiveness evaluation: a comparison of methods. *Am J Epidemiol* 2005;162:1016-23.

**10.** Azzato EM, Greenberg D, Shah M, Blows F, Driver KE, Caporaso NE*, et al.* Prevalent cases in observational studies of cancer survival: do they bias hazard ratio estimates? *Br J Cancer* 2009;100:1806-11.

**11.** Applebaum KM, Malloy EJ, Eisen EA. Left truncation, susceptibility, and bias in occupational cohort studies. *Epidemiology* 2011;22:599-606.

**12.** van Eekelen R, van Geloven N, van Wely M, Bhattacharya S, van der Veen F, Eijkemans MJ*, et al.* IVF for unexplained subfertility; whom should we treat? *Hum Reprod* 2019b;34:1249-59.

**13.** Vaupel JW, Manton KG, Stallard E. The impact of heterogeneity in individual frailty on the dynamics of mortality. *Demography* 1979;16:439-54.

**14.** Börnhorst C, Reinders T, Rathmann W, Bongaerts B, Haug U, Didelez V*, et al.* Avoiding Time-Related Biases: A Feasibility Study on Antidiabetic Drugs and Pancreatic Cancer Applying the Parametric g-Formula to a Large German Healthcare Database. *Clin Epidemiol* 2021;13:1027-38.

**15.** Eijkemans MJC, Leridon H, Keiding N, Slama R. A Systematic Comparison of Designs to Study Human Fecundity. *Epidemiology* 2019;30:120-29.

**16.** Austin PC. The use of propensity score methods with survival or time-to-event outcomes: reporting measures of effect similar to those used in randomized experiments. *Stat Med* 2014;33:1242-58.

**17.** Fireman B, Gruber S, Zhang Z, Wellman R, Nelson JC, Franklin J*, et al.* Consequences of Depletion of Susceptibles for Hazard Ratio Estimators Based on Propensity Scores. *Epidemiology* 2020;31:806-14.

**18.** Gran JM, Hoff R, Røysland K, Ledergerber B, Young J, Aalen OO. Estimating the treatment effect on the treated under time-dependent confounding in an application to the Swiss HIV Cohort Study. *Royal Stat Soc* 2018;67:103-25.

**19.** Hernán MA, Alonso A, Logan R, Grodstein F, Michels KB, Willett WC*, et al.* Observational studies analyzed like randomized experiments: an application to postmenopausal hormone therapy and coronary heart disease. *Epidemiology* 2008;19:766-79.

**20.** Harrell FE, Jr., Lee KL, Mark DB. Multivariable prognostic models: issues in developing models, evaluating assumptions and adequacy, and measuring and reducing errors. *Stat Med* 1996;15:361-87.

**21.** van Houwelingen H and Putter H. *Dynamic prediction in Clinical Survival Analysis. Monographs on statistics and applied probability*. Florida, United States: Chapman and Hall; 2012. 250 p.
**22.** Gran JM, Roysland K, Wolbers M, Didelez V, Sterne JA, Ledergerber B*, et al.* A sequential Cox approach for estimating the causal effect of treatment in the presence of time-dependent confounding applied to data from the Swiss HIV Cohort Study. *Stat Med* 2010;29:2757-68.
**23.** van Eekelen R, van Geloven N, van Wely M, McLernon DJ, Mol F, Custers IM*, et al.* Is IUI with ovarian stimulation effective in couples with unexplained subfertility? *Hum Reprod* 2019a;34:84-91.
**24.** Austin PC. Generating survival times to simulate Cox proportional hazards models with time-varying covariates. *Stat Med* 2012;31:3946-58.
**25.** Custers IM, Steures P, van der Steeg JW, van Dessel TJ, Bernardus RE, Bourdrez P*, et al.* External validation of a prediction model for an ongoing pregnancy after intrauterine insemination. *Fertil Steril* 2007;88:425-31.
**26.** van der Steeg JW, Steures P, Eijkemans MJ, Habbema JD, Hompes PG, Broekmans FJ*, et al.* Pregnancy is predictable: a large-scale prospective external validation of the prediction of spontaneous pregnancy in subfertile couples. *Hum Reprod* 2007;22:536-42.
**27.** van Eekelen R, Rosielle K, van Welie N, Dreyer K, van Wely M, Mol B*, et al.* Does the effectiveness of IUI in couples with unexplained subfertility depend on their prognosis of natural conception? A replication of the H2Oil study. *Human reproduction open* 2020;2020:hoaa047.
**28.** Van Eekelen R, Van Geloven N, Van Wely M, McLernon D, Mol F, Custers I*, et al.* Is IUI with ovarian stimulation effective in couples with unexplained subfertility? *Human Reproduction* 2019;34:84-91.
**29.** Wei L, Lin D, Weissfeld L. Regression analysis of multivariate incomplete failure time data by modeling marginal distributions. *Journal of the American Statistical Association* 1989;84:1065-73.
**30.** Team RC, *R: A language and environment for statistical computing. R Foundation for Statistical Computing, Vienna, Austria. [http://www.R-project.org/](http://www.R-project.org/).* 2017.
**31.** Luijken K, Spekreijse JJ, van Smeden M, Gardarsdottir H, Groenwold RHH. New-user and prevalent-user designs and the definition of study time origin in pharmacoepidemiology: A review of reporting practices. *Pharmacoepidemiol Drug Saf* 2021;30:960-74.
**32.** Lund JL, Richardson DB, Stürmer T. The active comparator, new user study design in pharmacoepidemiology: historical foundations and contemporary application. *Curr Epidemiol Rep* 2015;2:221-28.
**33.** Suissa S, Dell'Aniello S, Renoux C. The Prevalent New-user Design for Studies With no Active Comparator: The Example of Statins and Cancer. *Epidemiology* 2023;34:681-89.